\documentclass[onecolumn,secnumarabic,amssymb, nobibnotes, aps, prd]{revtex4-2}
\usepackage{amsmath}
\usepackage{amssymb,bbold}
\usepackage[usenames,dvipsnames]{color}
\usepackage{enumitem,kantlipsum}
\usepackage{lipsum}
\usepackage{natbib}
\usepackage{soul}
\usepackage{cancel}
\usepackage{dcolumn}
\usepackage{gensymb}
\usepackage{graphicx}
\usepackage{psfrag}
\usepackage{bm}
\usepackage[normalem]{ulem}
\usepackage{multirow}
\usepackage[colorlinks, citecolor = blue, urlcolor = blue]{hyperref}
\usepackage{cleveref}
\hypersetup{
  colorlinks=true,
  citecolor=blue,
  linkcolor=cyan,
  urlcolor=blue,
}

\newcommand{\BibitemShut}[1]{}
\def\beq{\begin{equation}}
\def\eeq{\end{equation}}
\def\bea{\begin{eqnarray}}
\def\eea{\end{eqnarray}}
\usepackage{hyperref}
\usepackage{epstopdf}
\AppendGraphicsExtensions{.pdf}

\begin{document}
\title{Dynamic structure factor of a driven-dissipative Bose-Hubbard model}
\author{Subhanka Mal\footnote{Authors contributed equally}}
\email{subhankamal@gmail.com}
\affiliation{School of Physical Sciences, Indian Association for the Cultivation of Science, Jadavpur, Kolkata 700032, India.}
\author{Anushree Dey\footnotemark[1]}
\email{anushree137@gmail.com}
\affiliation{School of Physical Sciences, Indian Association for the Cultivation of Science, Jadavpur, Kolkata 700032, India.}
\author{Kingshuk Adhikary}
\email{kingshuk804@gmail.com}
\address{Department of Optics, Palacký University, 17. Listopadu 1192/12, 779 00 Olomouc, Czech Republic}
\author{Bimalendu Deb}
\email{msbd@iacs.res.in}
\affiliation{School of Physical Sciences, Indian Association for the Cultivation of Science, Jadavpur, Kolkata 700032, India.}
\footnotetext{Authors contributed equally}

\begin{abstract}
Dynamic structure factor (DSF) is important for understanding excitations in many-body physics; it reveals  information about the spectral and spatial correlations of fluctuations in quantum systems. Collective phenomena like quantum phase transitions of ultracold atoms are addressed by harnessing density fluctuations. Here, we calculate the DSF of a nonequilibrium spinless Bose-Hubbard model (BHM) from the perspective of dissipative phase transition (DPT) in a steady state. Our methodology uses a homogeneous mean-field approximation to make the single-site hierarchy simpler and applies the Lindbladian perturbation
method (LPM) to go beyond the single site, limited by the ratio of the inter-site hopping term to the Liouvillian gap as a small parameter. 
 Our results show that the DSF near a DPT point is characteristically different from that away from the transition point, providing a clear density spectral signature of the DPT.  In addition to comparing the two numerical frameworks, the mean-field results serve as a benchmark for  proof-of-principle robustness of LPM. Despite the numerical difficulty, our methodology
provides a computationally accessible route for studying density fluctuations in an open lattice quantum system without requiring large-scale computation.

\end{abstract}
\maketitle

\section{Introduction}

Dynamic structure factor (DSF) provides important information about the temporal and spatial density fluctuations of a many-particle quantum system \cite{pines1966theory}.  In quantum mechanics, not all observables are measured at the same time; some, such as the DSF, require two-time correlations which are essential for probing dynamical responses. In contrast, the static structure factor captures equal-time density-density correlations and can be computed without explicit temporal information. The dynamical properties of density excitations,  are more realistically described by response functions that incorporate temporal separation.  These
 characteristics are essential in understanding how a system responds to perturbations over
 time, both in closed or open quantum systems. Because of its experimental accessibility,
 spatial density measurements are widely used to describe excitation spectra and critical
 behavior in a variety of platforms, including driven-dissipative quantum systems, photonic lattices, and ultracold atoms. Usually, DSF is calculated or measured   near an equilibrium thermodynamic or quantum phase of the system within the framework of linear response theory.  With the recent advent of  a nonequilibrium or driven dissipative phase transition (DPT)  \cite{PhysRevA.83.013611, PhysRevLett.105.015702, PhysRevA.93.033824, PhysRevA.95.013812, PhysRevLett.110.233601, PhysRevA.96.043809}, the question  arises how one can calculate DSF of a nonequilibrium system. 
 
 Among non-equilibrium platforms, driven-dissipative architectures have attracted considerable attention because of their experimental versatility in engineered systems that demonstrate dissipative phase transitions. Unlike their equilibrium counter
parts, these transitions occur in the steady states of open quantum systems and are  usually accompanied by dramatic changes in the systems' dynamical and spectral properties, as previously discussed \cite{Adhikary_PRA:2021}. Although, a variety of theoretical methods have been developed over the years to address nonequilibrium phases of a driven open quantum system \cite{weimer2015variational, dalibard1992wave, garrahan2010thermodynamics, diehl2010dynamical, prosen2008quantum, prosen2008third}, numerically calculating DSF of a driven dissipative lattice system is a challenging task.  Since the Lindbladian perturbation theory \cite{SCi_Rep_4_4887:2014,PRX_6_021037:2016} can systematically determine corrections to a mean-field density matrix of an open quantum system, it is perhaps worth applying this method to calculate the DSF of an open lattice system such as driven dissipative Bose-Hubbrad model. Perturbation theory is commonly applied to closed quantum systems. A variety of specialised perturbative methods have been employed in open quantum systems to study  finite-time evolution \cite{PRA_62_013819:2000}, full counting statistics \cite{PRL_100_150601:2008,PRB_82_155407:2010} and critical behavior \cite{PRL_110_195301:20013,arxiv_1309.7027:2013}. From an experimental point of view, the quasiparticle and collective excitations are measured by detecting the DSF of the system \cite{Steinhauer_PRL:2002}. The rapid advancement in cold-atom experimental techniques \cite{Weimer_NJP:2012} has made it possible  to detect the DSF of an equilibrium cold atomic system by using a weak probe field. In condensed matter systems, the response of the system is usually measured by neutron spectroscopy \cite{Sturm_ZNA:1993} or by scattering of radiation. In contrast, for cold atomic gases, stimulated light scattering or Bragg spectroscopy \cite{Stenger_PRL:1999} is used to calculate DSF and related quantities such as density response function \cite{Pino_PRA:2011, Landing_NatCom:2015} and collective excitations \cite{Hoinka_NatPhys:2017,Biss_PRL:2022}. The spectral function, which is related to the response of the system, has  been employed  to identify different phases of an equilibrium BHM \cite{Brown_NatPhys:2020,Bohrdt_PRL:2021}. Although DSF of various equilibrium lattice models has been extensively studied, that of a nonequilibrium lattice model undergoing DPT has not been significantly examined so far.  
 
In this work, we study the DSF of a driven dissipative Bose-Hubbard model (BHM), a paradigmatic setup that captures the interplay between interactions, coherent drive, tunnelling, and dissipation. Our focus is on how the
 structure factor encodes spectral signatures of a DPT and how these signatures evolve
 across the transition. To investigate this, we first compute the steady-state properties of
 the system within a single-site mean-field approximation, where a clear transition is observed as a function of the drive strength. This analysis is important as it has 
 already demonstrated how the Liouvillian gap behaves in the vicinity of DPT \cite{Adhikary_PRA:2021}. To go beyond mean-field, we apply Lindbladian perturbation theory using the ratio
 of the inter-site hopping strength to the Liouvillian gap as a small parameter. One can
 systematically take into account the effects of quantum fluctuations  which are very important near criticality. This study offers a theoretical framework and a computational approach to investigate non-equilibrium criticality
 through measurable spectral observables.  Our findings show that the DSF has distinct features near the DPT point, representing a
 spectral indicator of the underlying phase transition.  We show that, as 
 drive strength approaches the critical value, the half width at half maximum (HWHM) of DSF as a function frequency attains 
a minimum for all crystal momenta, and the peak of DSF appears at zero frequency, implying  
 that  the decay time of the density-density correlation function is the largest at the phase transition point.

The structure of the paper is as follows: In Sec.\ref{sec2} we present the model of driven dissipative BHM. We introduce the LPM and discuss its application to calculate density-density  correction functions of an open quantum system in Sec.\ref{sec2.1}. The description of DSF for the nonequilibrium BHM is introduced in Sec.\ref{sec2.2}. In Sec.\ref{sec3} we illustrate our numerical results. The results for the mean-field description are analysed in Sec.\ref{sec3.1}. The numerical analysis of the DSF is illustrated in Sec.\ref{sec3.2}. Finally, in Sec.\ref{sec4}, the conclusions of this paper are presented.

\section{The model}\label{sec2}

The Hamiltonian of a driven BHM  is given by
\begin{eqnarray}
H=-\frac{J}{z}\sum_{\langle j,k\rangle}(\hat a^\dagger_j\hat a_k + h.c.) + \frac{U}{2}\sum_j \hat a^\dagger_j\hat a^\dagger_j\hat a_j\hat a_j +\epsilon\sum_j \hat a^\dagger_j\hat a_j +H_{drive}
\label{eq1}
\end{eqnarray}
where $\hat a^\dagger_j$ and $\hat a_j$ are bosonic creation and annihilation operators for the $j$-site respectively, $J (>0)$ is the tunnelling strength, $z$ is the coordination number for each site. $\epsilon$ represents the energy per particle in each site. The drive term in the Hamiltonian is given by,
\begin{eqnarray}
H_{drive} = \sum_j F\hat a^\dagger_j e^{-i\omega t}+F^* \hat a_j e^{i\omega t}
\label{eq2}
\end{eqnarray}
where $F$ and frequency $\omega$ are the strength and frequency of the drive, respectively. In a frame that rotates with the frequency $\omega$, the effective Hamiltonian can be rewritten as 
\begin{eqnarray}
H=-\frac{J}{z}\sum_{\langle j,k\rangle}(\hat a^\dagger_j\hat a_k + h.c.) - \hbar\Delta\omega\sum_j \hat a^\dagger_j\hat a_j +\sum_j (F\hat a^\dagger_j + F^*\hat a_j) + \frac{U}{2} \sum_j \hat a^\dagger_j \hat a^\dagger_j \hat a_j \hat a_j \nonumber\\
\label{eq3}
\end{eqnarray}
where $\Delta\omega = \omega - \epsilon/\hbar$ is the detuning. In the presence of system-bath coupling the evolution of the system becomes non-unitary. One can construct the density matrix of the system governed by the Lindblad master equation \cite{lindblad1976generators,lindblad1979non}
\begin{eqnarray}
\dot{\rho}(t) &=& -i[H,\rho(t)] + \sum_j\gamma_j\mathcal{D}[\hat a_j]\rho(t)\nonumber\\
&=& \mathcal{L}\rho(t)
\label{eq4}
\end{eqnarray}
where $\rho$ is the density matrix of the system and $\gamma_j$ is the rate of damping from $j$-th site. The dissipation of the system is described by the super-operator \cite{lindblad1979non}
\begin{eqnarray}
\mathcal{D}[\hat a_j]\rho = \hat a_j\rho\hat a_j^\dagger - \frac{1}{2}\{\hat a_j^\dagger \hat a_j,\rho\}
\label{eq5}
\end{eqnarray}
and $\mathcal{L}$ is the Liouville super-operator. 

\begin{figure}
    \centering
    \includegraphics[width=0.8\linewidth]{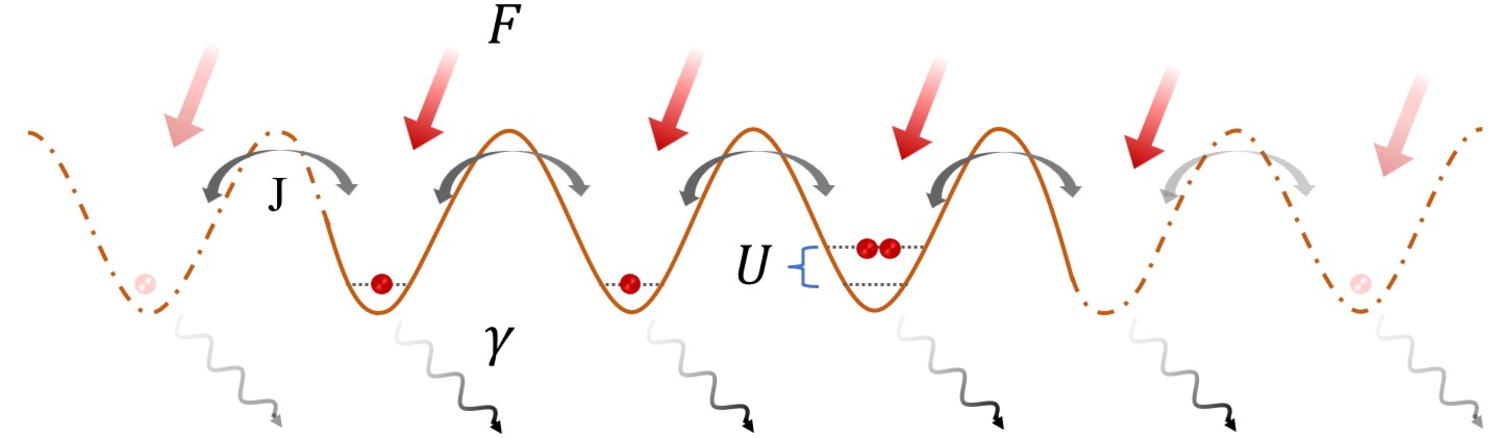}
    \caption{Schematic diagram of a one-dimensional Bose-Hubbard model with a drive with strength $F$ and dissipation coefficient $\gamma$ for each site. $J$ represents the inter-site tunnelling coefficient. The on-site interaction between the two particles is $U$.}
    \label{fig:schematic}
\end{figure}

A single-site mean-field description of this driven dissipative BHM can be developed by decoupling approximation \cite{Sheshadri_EPL:1993}
\begin{eqnarray}
\hat a_j^\dagger\hat a_k &\approx& \langle \hat a_j^\dagger\rangle \hat a_k + \hat a_j^\dagger \langle\hat a_k\rangle - \langle \hat a_j^\dagger\rangle \langle\hat a_k\rangle \nonumber\\
&=& \psi^*\hat a_k +\psi \hat a_j^\dagger -|\psi|^2
\label{eq5.1}
\end{eqnarray}
which reduces the multi-site problem into a single-site problem. Here, the order parameter $\psi = \langle\hat a_j\rangle$ is site independent. Thus, the average number of particles per site is $\bar n = \langle \hat a_j^\dagger\hat a_j\rangle$ while the coherent number density is given by $n_{co} = |\psi|^2$. According to the earlier studies \cite{Boite_PRA:2014,Adhikary_PRA:2021}, this approximation holds good both in equilibrium and nonequilibrium BHM to describe zero-momentum states. However, the mean-field approach is incapable of capturing the effects of finite momentum states, which may lead to excitations in the system.

Now, we discuss in some detail the possible realisations of our model.  In recent times,  a variety of experimental platforms capable of realising such systems under controlled laboratory conditions have been developed.  Among such systems, the most notable ones are driven cavity and circuit quantum electrodynamics (QED) architectures.  Currently, there is significant interest in a system of cavity arrays where the tunnelling of photons across adjacent cavities leads to a wide range of nonequilibrium phenomena \cite{Hartmann_LPR:2008, Tomadin_JOSAB:2010, Houck_NatPhys:2012, Carusotto_RMP:2013, Schmidt_AP:2013, Underwood_PRA:2012, Toyoda_PRL:2013, Raftery_PRX:2014}.  Recent progress in photonic crystals \cite{armani2003ultra}, optical microcavities \cite{bayindir2000tight}, and superconducting devices \cite{wallraff2004strong} has led to a class of coupled-cavity models that offer a solution to the issue of individual driving. In contrast, it is difficult to drive individual sites of an optical lattice, since the inter-site optical lattice spacing  is typically of the size of half an optical  wavelength. An optical lattice has been widely employed as a unique platform for studying quantum phase transitions, particularly  in equilibrium Bose-Hubbard model \cite{Jaksch_PRL:1998,Greiner_Nat:2002,Roth_JPB:2004,Roux_NJP:2013}. The dynamics of several interacting photons in driven nonequilibrium cavity arrays \cite{carusotto2009fermionized, tomadin2010signatures, hartmann2010polariton, nunnenkamp2011synthetic, nissen2012nonequilibrium, grujic2012non, grujic2013repulsively, jin2013photon, le2013steady, vivek2023nonequilibrium} leads to the emergent quantum phenomena \cite{diehl2008quantum,diehl2010dynamical,baumann2010dicke,lesanovsky2010thermalization,lee2011antiferromagnetic,ludwig2013quantum} and nonequilibrium photonic phases \cite{PhysRevA.96.023839}. Although, our model pertains to a general dissipative BHM, this can be more suitably fit into  an array of nonlinear coupled cavities \cite{PhysRevB.78.075320}.

Before we discuss the Lindbladian perturbation method (LPM), it is worthwhile to comment on some numerical techniques commonly used to study lattice models.  Although, some advanced numerical techniques, such as matrix-product algorithm \cite{PRL_93_207204:2004, PRL_93_207205:2004}, self-consistent projection operator method \cite{PRB_89_245108:2014}, Lanczos-type algorithm \cite{Roth_JPB:2004}, transverse-field Ising model \cite{Baez_PNAS:2020} and recently proposed quantum approximate optimization algorithm \cite{Pagano_PNAS:2025} can handle moderate-sized lattice systems in equilibrium cases, each has its own
 set of limitations.  On the other hand, for a nonequilibrium lattice model such as driven dissipative BHM, numerical methods of 
  solving the master equation directly such as  exact diagonalisation method \cite{PRA_47_3311:1993, PRA_89_052133:2014}, averaging of quantum trajectories \cite{RMP_70_101:1998}, etc. are very hard to implement on a computer as the size of the lattice grows. The rapid growth of the Hilbert space with the size of the lattice makes it difficult to numerically calculate the density matrix of an open quantum system of an $N$-site lattice with $N \ge 4$.  

\subsection{The Lindbladian perturbation method }\label{sec2.1}

 The Liouvillian superoperator $\mathcal{L}$ is a non-Hermitian operator and the left and right $\mu$-th eigenmatrices can be represented as \cite{Li_SciRep:2014}
\begin{eqnarray}
    \mathcal{L}u_\mu = \lambda_\mu u_\mu, \hspace{0.2in} \mathcal{L}^\dagger w_\mu = \lambda_\mu^*w_\mu
    \label{eq6}
\end{eqnarray}
These eigenmatrices follow the orthonormalization condition $\langle w_\mu,u_\nu\rangle = {\rm Tr}[w_\mu^\dagger u_\nu] = \delta_{\mu\nu}$. The density matrix perturbation theory is applicable if the total Liouvillian operator can be represented in the following form
\begin{eqnarray}
\mathcal{L} = \mathcal{L}_0 + \alpha\mathcal{L}_1
\label{eq7}
\end{eqnarray}
where $\alpha$ is a small parameter, $\mathcal{L}_0$ is the unperturbed superoperator and $\mathcal{L}_1$ is the perturbative correction term. Therefore,
\begin{eqnarray}
\mathcal{L}_0 u_\mu = \lambda_\mu^{(0)} u_\mu^{(0)} \hspace{.1 in} {\rm and} \hspace{.1 in} \mathcal{L}_0^\dagger w_\mu = \lambda_\mu^{(0)*} w_\mu^{(0)}  \nonumber
\end{eqnarray}
and the steady-state density matrix is identified as the eigenmatrix of the Liouvillian with zero eigenvalue, i.e., $\mathcal{L}\rho_{ss} = 0$ (where $\rho_{ss}$ is the steady-state density matrix). For this case, both $\lambda_\mu$ as well as $\lambda_\mu^{(0)}$ vanish.

Expanding the eigenvalues and eigenmatrices in $\alpha$ we have,
\begin{eqnarray}
\lambda_\mu = \sum_{j=0}^\infty \alpha^j \lambda_\mu^{(j)}, \hspace{0.2 in} u_\mu = \sum_{j=0}^\infty \alpha^j u_\mu^{(j)}
\label{eq8}
\end{eqnarray}
The index $j$ represents the order of perturbation. Using Eqs. (\ref{eq6}) and (\ref{eq8}) one obtains the recursive relation
\begin{eqnarray}
\left(\mathcal{L}_0 - \lambda_\mu^{(0)}\right)u_\mu^{(j)} = -\mathcal{L}_1u_\mu^{(j-1)} + \sum_{k=1}^j \lambda_\mu^{(k)}u_\mu^{(j-k)}
\label{eq9}
\end{eqnarray}
From Eq. (\ref{eq9}), it is straightforward to calculate the perturbative corrections of the eigenmatrices. However, as $\mathcal{L}_0$ is non-Hermitian, the usual inversion method does not work here. Therefore, we follow the generalised inversion technique of any singular matrix, namely the Moore-Penrose pseudoinverse formalism \cite{moore1920reciprocal,penrose1955generalized} to calculate the correction terms:
\begin{eqnarray}
    u_\mu^{(j)} = \left(\mathcal{L}_0-\lambda_\mu^{(0)}\right)^{\leftharpoonup1} {\Big(}-\mathcal{L}_1u_\mu^{(j-1)} + \sum_{k=1}^j \lambda_\mu^{(k)}u_\mu^{(j-k)} {\Big)}
    \label{eq10}
\end{eqnarray}

The steady-state density matrix is the right eigenmatrix of the Liouvillian superoperator corresponding to the zero eigenvalue, i.e.,
\begin{eqnarray}
    \mathcal{L}\rho_{ss} = \mathcal{L}u_0 = 0
    \label{eq11}
\end{eqnarray}
and since $\mathcal{L}_0\rho_{ss}^{(0)} = 0$, one must have
\begin{eqnarray}
    \lambda_\mu^{(j)} = 0
    \label{eq12}
\end{eqnarray}
for all $j$. Combining Eqs. (\ref{eq10}) and (\ref{eq12}) we have the simplified form for the perturbative correction of the steady state density matrix
\begin{eqnarray}
    \rho_{ss}^{(j)} = -\mathcal{L}_0^{\leftharpoonup1}\mathcal{L}_1\rho_{ss}^{(j-1)}
    \label{eq13}
\end{eqnarray}
The corresponding total Liouvillian is given by Eq. (\ref{eq4}).

\subsection{Dynamic structure factor of a nonequilibrium BHM}\label{sec2.2}
Now , we discuss the application of this method to our model. To apply density matrix perturbation theory, we choose the tunnel coupling $J$ to be small compared to other parameters of the system. We construct the unperturbed Liouvillian from the reduced single-site Hamiltonian and take the rest of the terms as perturbation. So
\begin{eqnarray}
    \mathcal{L}_0\rho = \sum_j {\Big\{}-i{\Big[}{\Big(}\beta^*\hat{a}_j+\beta\hat{a}_j^\dagger - \Delta\omega\hat{a}_j^\dagger\hat{a}_j + \frac{U}{2}\hat{a}_j^{\dagger2}\hat{a}_j^2 + \frac{J}{2}|\psi|^2{\Big)},\rho{\Big]} + \gamma\mathcal{D}[\hat{a}_j]\rho{\Big\}} \nonumber \\
    \label{eq14}
\end{eqnarray}
and 
\begin{eqnarray}
    \alpha\mathcal{L}_1\rho = J\sum_j i{\Big[(}(\hat{a}_j^\dagger\hat{a}_{j+1} + c.c.) -\psi(\hat{a}_j^\dagger+\hat{a}_j) + |\psi|^2{\Big)},\rho{\Big]}
    \label{eq15}
\end{eqnarray}
where $\psi = \langle\hat{a}_j\rangle = \langle\hat{a}_{j+1}\rangle$ calculated in the mean-field approach, $\beta = F-\psi J$. We set $z=1$ as we consider an one-dimensional Bose-Hubbard chain and $\hbar = 1$. The first order correction to the steady state is calculated as
\begin{eqnarray}
    \rho_{ss}^{(1)} = -\mathcal{L}_0^{\leftharpoonup1}\mathcal{L}_1\rho_{ss}^{(0)}
    \label{eq16}
\end{eqnarray}
$\rho_{ss}^{(0)}$ is the steady state density matrix calculated in the mean-field approximation (MFA). The resultant steady state after the first order correction is 
\begin{eqnarray}
    \rho_{ss} = \mathcal{N}\left[\rho_{ss}^{(0)} + \alpha \rho_{ss}^{(1)}\right]\\
    \label{eq17}
\end{eqnarray}
where $\alpha = J$ and $\mathcal{N}[\hat{O}] = \hat{O}/{\rm Tr}[\hat{O}]$.

For a one dimensional $N_s$-site system, the annihilation operator in momentum space is written as
\begin{eqnarray}
\hat c_k = \frac{1}{\sqrt{N_s}}\sum_{j=1}^{N_s} e^{i(2\pi j/N_s)k}\hat a_j
\end{eqnarray}  
The density operator 
\begin{eqnarray}
\hat \rho_q = \sum_k \hat c_{k+q}^\dagger \hat c_k
\end{eqnarray}
The density-density correlation function
\begin{eqnarray}
\chi (q, \tau-\tau') = -\langle T_\tau[\hat\rho_q(\tau)\hat\rho_{-q}(\tau')]\rangle
\end{eqnarray}
where $T_\tau$ is the time-ordering operator. The corresponding DSF is given by the Fourier transform for the correlation function
\begin{eqnarray}
S(\omega,k) = \frac{1}{N_s} \sum_{j=1}^{N_s}\int \chi(r_j,t)e^{i\omega t} dt
\end{eqnarray}
where $\chi(r_i-r_j,\tau-\tau') = -iT_\tau[\langle \hat n_i(\tau)\hat n_j(\tau')\rangle]$ is the real space density correlation with $\hat n_i =\hat a_i^\dagger \hat a_i$ being the density of $i$-th site.

\section{Results and discussions}\label{sec3}

Now we present the results. At first we describe the mean-field results obtained by decoupling the model Hamiltonian (\ref{eq4}) in single-site decomposition (as shown in Eq. (\ref{eq5.1})) for a small tunnel coupling $J/\gamma=0.1$ in the scale of damping coefficient, $\gamma$ which is similar for all sites. Next, we compare the mean-field results with the results obtained in the beyond-mean-field perturbative approach. We mainly focus on the behavior of average number of particles in each site and density-density fluctuations as a function of drive strength for different crystal momenta. By continuously adjusting the driving strength for fixed values of other system parameters, we obtain a nonequilibrium quantum phase transition indicated by the sudden jump in average particle number per site and also by critical behavior near the phase transition point \cite{Adhikary_PRA:2021}. In our numerical calculations we first check the convergence of our results by changing the number of Fock bases per site. For the case of small driving strength, the minimum number of Fock bases per site $N_F$ required for a convergent result remains small ($\le 6$). However, as the driving strength increases and approaches a DPT, one has to increase the value of $N_F$ to have convergent results. 

\subsection{Mean-field analysis}\label{sec3.1}
We first analyse the results in mean-field approximation for small tunnel coupling $J/\gamma = 0.1$, interaction strength $U = \gamma$ and detuning $\Delta\omega = 2\gamma$. For this case we have considered $N_F = 10$ to obtain convergent results for the entire range of driving strength for our chosen system parameters. Since the mean-field gives the single-site description, the dimension of the density matrix is $10\times10$ and the associated Lindbladian has the dimension of $10^2 \times 10^2$. We make use of the QuTip toolbox \cite{johansson2012qutip} available in Python library to solve for the steady-state density matrix.  
In the steady-state limit, we calculate the average photon number per site $\bar n = \langle \hat a^\dagger\hat a\rangle$, the coherent part of the density $n_{co} = |\psi^2|$ and the zero-time second-order correlation function $g^{(2)}(0) = \frac{\langle\hat a^\dagger\hat a^\dagger\hat a\hat a\rangle}{\langle\hat a^\dagger\hat a\rangle^2}=\frac{\langle\hat{n}^2\rangle-\langle \hat n\rangle}{\langle \hat n\rangle^2}$. So the number variance is $(\Delta n)^2=\langle\hat{n}^2\rangle-\langle \hat n\rangle^2=(g^{(2)}(0)-1)\langle\hat{n}\rangle^2+\langle\hat{n}\rangle$.

\begin{figure}[h]
   \centering
   \includegraphics[width=\linewidth]{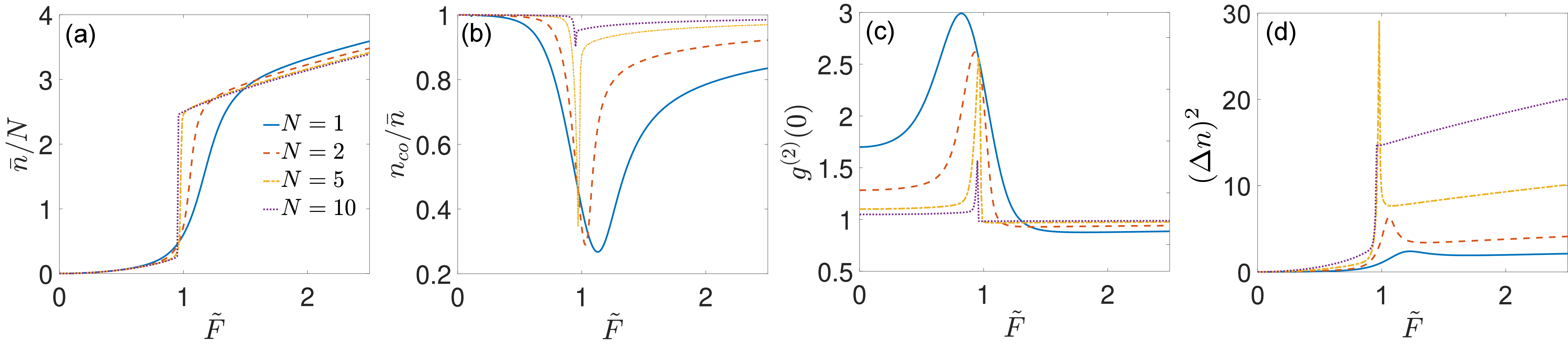}
   \caption{(a) Scaled photon density $\bar{n}/N$, (b) $n_{co}/\bar n$, (c) the second-order correlation function $g^{(2)}(0)$ and (d) the number variance as a function of the scaled drive amplitude $\tilde{F}$ for $\Delta \omega= 2\gamma$, $\tilde{U }= \gamma$ and $J/\gamma=0.1$ .}
   \label{fig:J=0.1}
\end{figure}

To capture the critical behavior of the system at or near the critical drive strength \cite{Adhikary_PRA:2021}, we introduce the dimensionless parameter $N$ such that $U=\tilde{U}/N$ and $F=\sqrt{N}\tilde{F}$ and we will consider the limit $N\rightarrow\infty$ such that $U F^2$ remains constant. This is an analogue of the thermodynamic limit for the driven system. In Fig.\ref{fig:J=0.1}(a) and (b) the scaled average particle number $\bar n/N$ per site and corresponding superfluid fraction $n_{co}/\bar{n}$ are plotted as a function of $\tilde{F}$ for $\Delta \omega= 2\gamma$, $\tilde{U}= \gamma$ and $J/\gamma=0.1$ for increasing values of $N$. For lower values of $N$, $\bar n/N$ gradually increases from $\tilde{F}=0.5$. It is observed that as $N$ increases, the variation of $\tilde{n}/N$ becomes sharper. As $N\rightarrow\infty$ ($N=10$ for our case), we observe a discontinuous jump of $\bar{n}/N$ at $\tilde{F}=0.95$ suggesting a first-order nonequilibrium phase transition (NEPT).  This is the nonequilibrium counterpart of the thermodynamic limit.
 The superfluid fraction $n_{co}/\bar{n}$ also shows similar behavior around the phase transition point. Moreover, $g^{(2)}(0)$ and $(\Delta n)^2$ in Fig.\ref{fig:J=0.1}(c) and (d) exhibit strongly discontinuous or nonanalytic behavior at the transition point, revealing large number fluctuations near the critical point. Another way of detecting this critical point is to calculate critical exponent slightly away from the critical point, and to show the power law behavior. In an earlier study \cite{Adhikary_PRA:2021}, we have found the critical exponent to be independent of the parameter $N$.  
\begin{figure}[h]
    \centering
     \includegraphics[width=0.75\linewidth]{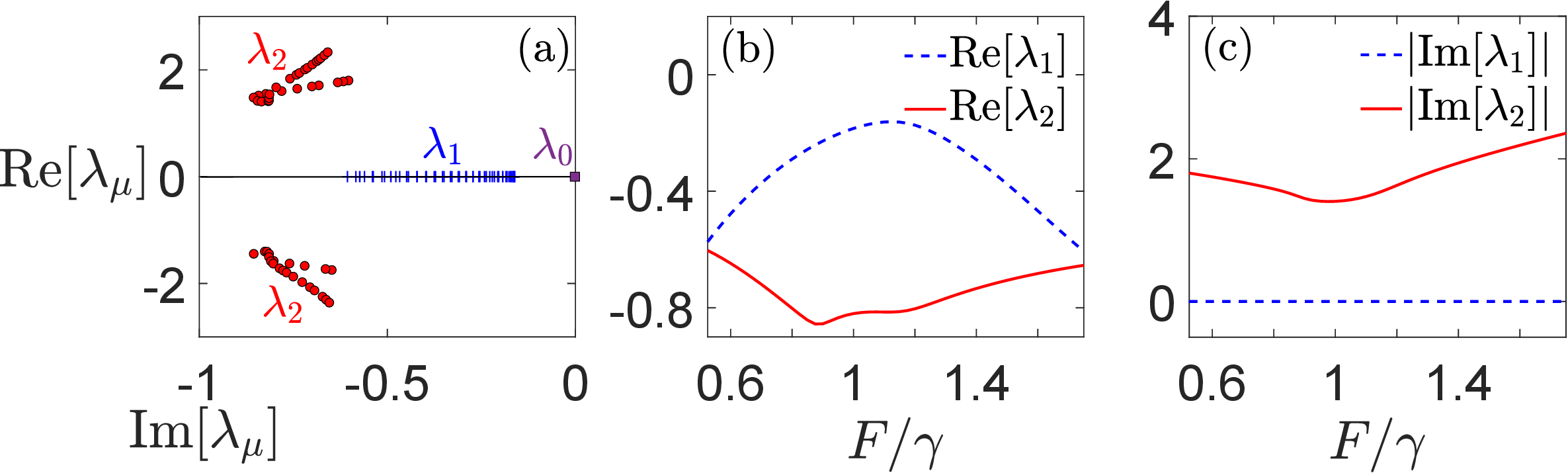}
    \caption{Eigenvalue spectrum of the Liouvillian for a range of drive strengths near the dissipative phase transition under mean-field approximation. (a) The lowest three eigenvalues $\lambda_\mu$ are shown in the complex plane for the drive strength in the range between $0.5\leq F/\gamma\leq1.75$. The real (b) and imaginary (c) parts of $\lambda_1$ and $\lambda_2$ are shown as a function of $F/\gamma$. The other system parameters are similar as in Fig.\ref{fig:J=0.1}.}
    \label{fig:lam1_lam2}
\end{figure}

We show the eigenvalue spectrum of the lowest three eigenvalues (in terms of the real parts of the eigenvalues) of the Liouvillian for a range of drive strengths between $0.5 \le F/\gamma \le 1.75$ with $\Delta\omega = 2\gamma$, $U/\gamma = 1.0$ and $J/\gamma = 0.1$ in Fig.\ref{fig:lam1_lam2}. We arrange the eigenvalues in the ascending order in terms of their absolute magnitudes. The lowest eigenvalue $\lambda_0$, associated with the steady-state density matrix, is always zero, as shown in Fig.\ref{fig:lam1_lam2}(a). On the other hand, while the second eigenvalue $\lambda_1$ is real, the third eigenvalue is always complex in this range. This is explicitly demonstrated in Fig.\ref{fig:lam1_lam2}(b) and (c). It is important to note that, $\lambda_1$ displays a maximum near the phase transition point $F/\gamma = 0.95$. This feature is known as the `gap closing' near a dissipative phase transition. 
The maximum value $|\lambda_1|/\gamma$ near the transition point is $0.16$ which is still larger than that $J/\gamma=0.1$ considered for our model calculations. Therefore, the results of the perturbation theory near the transition point may not be quantitatively accurate enough since $\frac{J}{\lambda_1}\approx 0.6$ near the transition point.

\subsection{Dynamic structure factor}\label{sec3.2}
We next obtain the Lindbladian perturbative results by solving Eqs.(\ref{eq14}), (\ref{eq15}), (\ref{eq16}) and (\ref{eq17}). Here, we consider a three-site Hamiltonian with periodic boundary conditions and check the convergence of our results for small hopping term $J = 0.1\gamma$. For our calculations, we set $U = \gamma$ and detuning $\Delta\omega = 2\gamma$. For the three-site system we consider $N_F = 5$ and the corresponding density matrix of the system becomes $5^3 \times 5^3$. Thus, the associated Lindbladian becomes $5^6\times5^6$ dimensional. These large dimensional quantities have made the calculations computationally challenging. Further increase in number of Fock bases per site needs additional symmetries to reduce the computational cost and is beyond the scope of this study. Nevertheless, we obtain convergent results for $N_F = 5$ for the range of driving strength $0\leq F \leq 1$ with the system parameters as mentioned above. 
\begin{figure}[h]
    \centering
     \includegraphics[width=0.75\linewidth]{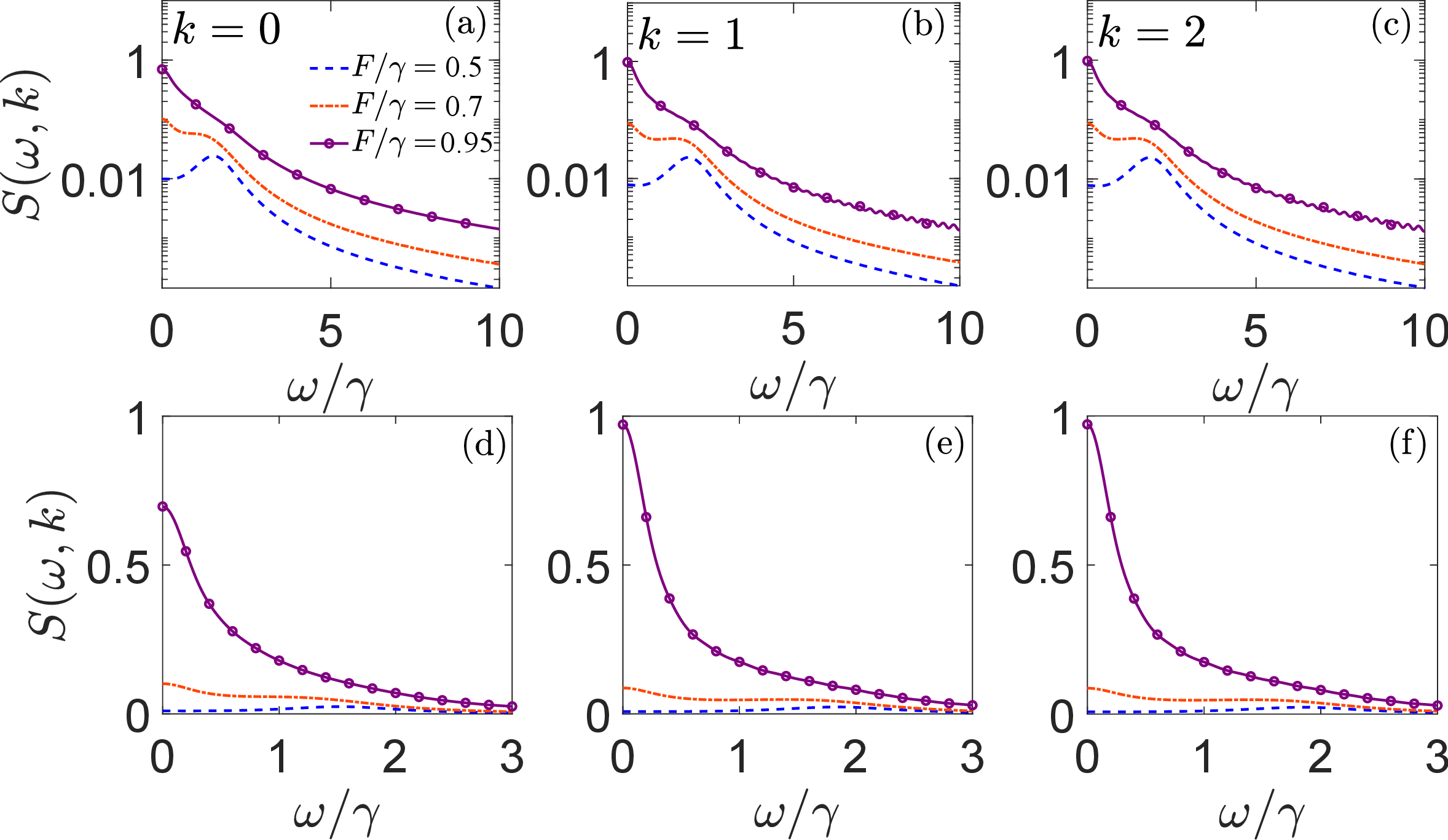}
    \caption{Dynamic structure factor $S(\omega,k)$ as a function of frequency $\omega$ for discrete wave numbers $k = 0$ (a, d), $1$ (b, e), $2$ (c, f) and for different values of drive strength $F$. The upper panel (a, b, c) is shown in semilogarithmic scale, while the lower panel (d, e, f) is shown in linear scale. $S(\omega,k)$ is shown for $F/\gamma = 0.5$ (blue dashed line);  $F/\gamma = 0.7$ (orange dash-dotted line) and  $F/\gamma = 0.95$ (purple line in circular marker).}
    \label{fig:structure_surf_line}
\end{figure}
\begin{figure}[h]
    \centering
     \includegraphics[width=0.8\linewidth]{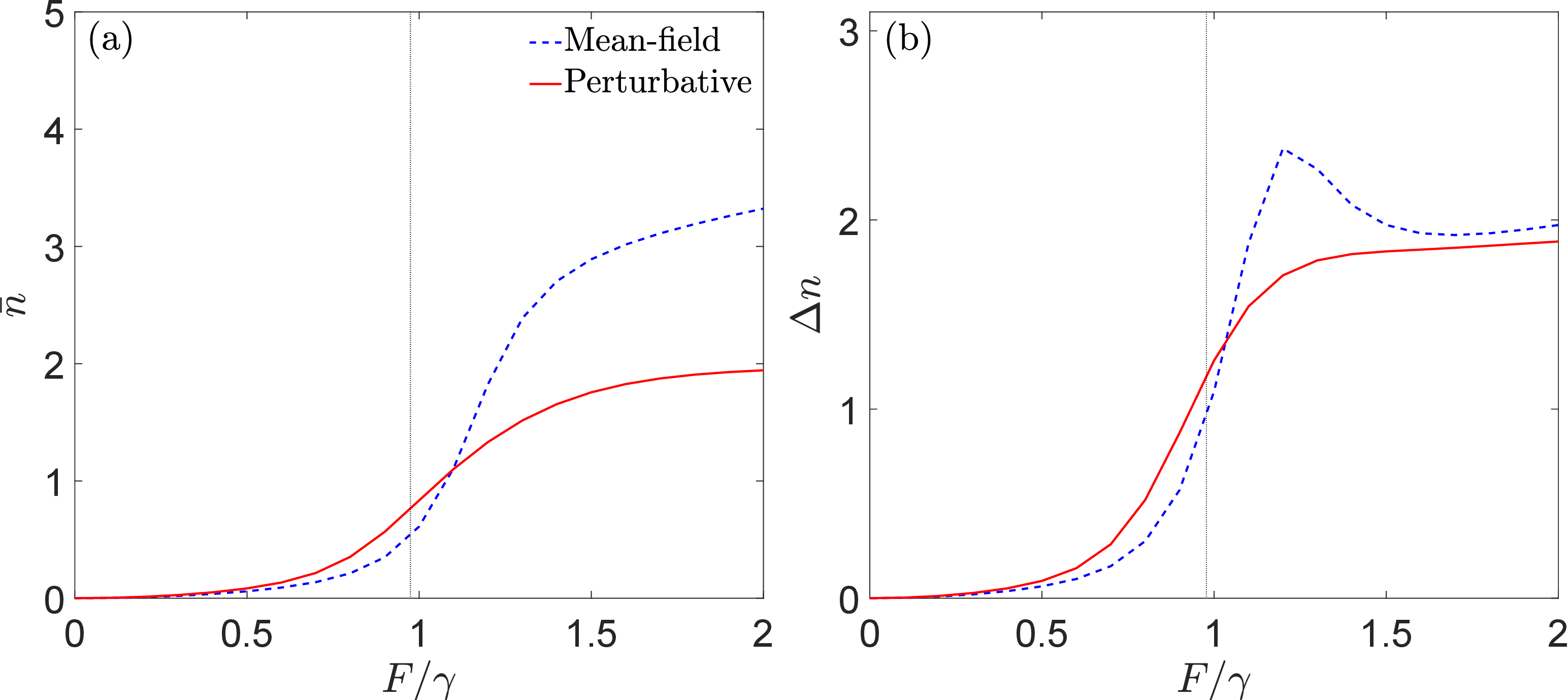}
\caption{Variation of average population (a) and number fluctuation (b) in each site at steady-state as a function of dimensionless drive strength $F/\gamma$ for the dimensionless parameters $\Delta\omega/\gamma = 2.0$, $J/\gamma = 0.1$ and $U/\gamma = 1.0$. The blue dashed line corresponds to the mean-field theory, while the red solid line corresponds to the perturbation theory. The dotted grey line marks the DPT point at $F/\gamma = 0.95$.}
\label{fig:comparison}
\end{figure}

In Fig.\ref{fig:structure_surf_line}, we show the structure factor $S(\omega,k)$ as a function of dimensionless frequency $\omega/\gamma$ for three discrete values of wave number $k=0$ (a, d), $k=1$ (b, e) and $k = 2$ (c, f). The lattice constant `$a$' is taken to be unity. For small drive strength, $F/\gamma = 0.5$ the spectrum of the structure factor has a single peak at finite $\omega/\gamma$. For an intermediate value of drive strength, i.e., for $F/\gamma = 0.7$, the side-peak starts to vanish. It is important to note that the width of the spectrum decreases and the peak value increases with increasing drive strength. Finally, as the dimensionless drive strength $F/\gamma$ approaches the critical value, the width of the spectrum  becomes minimum with the peak  appearing at $\omega = 0$ as a  signature of the dissipative phase transition. This means that, in time domain, the decay time of the density-density correlation function is the largest at the phase transition point. The fact that at the phase transition point the side peak vanishes implies that while the decay of the correlation function can be oscillatory for parameters away from the critical point, at the critical point the decay is non-oscillatory.  One can notice that, at the transition point, the peak value increases with increasing $k$.  It is also to be noted that, at the transition point, the widths of $S(\omega,k)$ for different $k$ values are same. This behavior qualitatively signifies that, near a phase transition point, both $k = 0$ and $k >0$ parts of the density fluctuations contribute equally to the DSF. The half width at half maximum  of the spectrum at the phase transition point  is equal to the nonzero real eigenvalue of the Liouvillian operator closest to zero,  known as Liouvillian gap. We find that, at the transition point $F/\gamma = 0.95$, the HWHM is $ 0.25\gamma$ which is found to be equal  to  the Liouvillian gap of the system with perturbative corrections. Note that the value of the Liouvillian gap at DPT is $0.18\gamma$ for the mean-field steady-state density matrix, which is considerably smaller than the LPM result.  It is to be further noted that the DSF fulfils the sum rule that relates it with the static structure factor $S(k) = S(\omega=0, k)$, namely, the relation $\int d\omega S(\omega, k) = N_s S(k)$.  

 
Finally, we make a few comments about the validity regime of LPM as used in our model by comparing the mean-field results with those  obtained from LPM. We show the variation of average occupation $\bar n$ as a function of dimensionless drive strength $F/\gamma$ in the steady state in Fig.\ref{fig:comparison}. The average population for all three sites are equal, and we set $\bar n_i = \bar n$ for all three values of $i$. For small driving strength, the mean-field results agree quite well with the perturbative ones. However, as the driving strength $F/\gamma$ approaches the phase transition point $F/\gamma = 0.95$, the two results start to differ since the first order perturbative approximation at this point is not good enough ($\frac{J}{|\lambda_1|}\approx 0.55$).

\section{Conclusions}\label{sec4}

In conclusion, we have calculated DSF $S(\omega, k)$ of a driven dissipative Bose-Hubbard model by using Lindbladian perturbation method and demonstrated the density spectral signature of dissipative phase transition. The minimum HWHM of the DSF corresponds to the DPT resulting in a sharp peak structure with its peak value appearing at $\omega = 0$ for all $k$. The inclusion of small perturbation in the form of tunnel coupling enables capturing the (discrete) finite lattice momentum that is neglected in the mean-field description. Our study also shows that although the contribution of finite momentum is small away from the DPT, at the phase transition point, the finite momentum contribution to the density correlation is significantly strong. As noted above, our model can be realised with an array of driven cavities, and DSF may be detected through cavity transmission. A more generalised detection technique of $S(\omega, k)$ for a Bose-Hubbard model has recently been proposed through weak measurement \cite{Altuntas_arxiv:2024} and is subject to experimental verification. Extension of this model to multicomponent bosonic and fermionic driven dissipative systems may remarkably pave the path to identify the dynamical properties near a DPT. Finally, our analysis can be applied under suitable approximations to enhance the system size for better understanding the quantum nature.

\section*{References}
\bibliographystyle{apsrev4-2}
\bibliography{cite_liou}
\end{document}